\documentstyle[12pt,epsfig]{article}
\advance\textheight by \topskip
\begin{document}
\tolerance=100000
\thispagestyle{empty}
\setcounter{page}{0}

\newcommand{\be}{\begin{equation}}
\newcommand{\ee}{\end{equation}}
\newcommand{\br}{\begin{eqnarray}}
\newcommand{\er}{\end{eqnarray}}
\newcommand{\ba}{\begin{array}}
\newcommand{\ea}{\end{array}}
\newcommand{\bi}{\begin{itemize}}
\newcommand{\ei}{\end{itemize}}
\newcommand{\bn}{\begin{enumerate}}
\newcommand{\en}{\end{enumerate}}
\newcommand{\bc}{\begin{center}}
\newcommand{\ec}{\end{center}}
\newcommand{\ul}{\underline}
\newcommand{\ol}{\overline}
\newcommand{\ar}{\rightarrow}
\newcommand{\sm}{${\cal {SM}}$}
\newcommand{\mssm}{${\cal {MSSM}}$}
\newcommand{\susy}{{{SUSY}}}
\def\epem{\ifmmode{e^+ e^-} \else{$e^+ e^-$} \fi}
\newcommand{\Dir}{\kern -6.4pt\Big{/}}
\newcommand{\Dirin}{\kern -10.4pt\Big{/}\kern 4.4pt}
\newcommand{\DDir}{\kern -7.6pt\Big{/}}
\newcommand{\DGir}{\kern -6.0pt\Big{/}}
\newcommand{\eett}{$e^+e^-\rightarrow t\bar t $}
\newcommand{\eeZphi}{$e^+e^-\rightarrow Z\phi $}
\newcommand{\eeZH}{$e^+e^-\rightarrow ZH $}
\newcommand{\eeAH}{$e^+e^-\rightarrow AH $}
\newcommand{\eehww}{$e^+e^-\rightarrow hW^+W^-$}
\newcommand{\bbww}{$b\bar b W^+W^-$}
\newcommand{\eebbww}{$e^+e^-\rightarrow b\bar b W^+W^-$}
\newcommand{\ttbbww}{$t\bar t\rightarrow b\bar b W^+W^-$}
\newcommand{\eettbbww}{$e^+e^-\rightarrow t\bar t\rightarrow b\bar b W^+W^-$}
\newcommand{\bbb}{$ b\bar b$}
\newcommand{\ttb}{$ t\bar t$}
\def\Ord{\buildrel{\scriptscriptstyle <}\over{\scriptscriptstyle\sim}}
%
\def\OOrd{\buildrel{\scriptscriptstyle >}\over{\scriptscriptstyle\sim}}
\def\Ord{\buildrel{\scriptscriptstyle <}\over{\scriptscriptstyle\sim}}
\def\OOrd{\buildrel{\scriptscriptstyle >}\over{\scriptscriptstyle\sim}}
\def\pl #1 #2 #3 {{\it Phys.~Lett.} {\bf#1} (#2) #3}
\def\np #1 #2 #3 {{\it Nucl.~Phys.} {\bf#1} (#2) #3}
\def\zp #1 #2 #3 {{\it Z.~Phys.} {\bf#1} (#2) #3}
\def\pr #1 #2 #3 {{\it Phys.~Rev.} {\bf#1} (#2) #3}
\def\prep #1 #2 #3 {{\it Phys.~Rep.} {\bf#1} (#2) #3}
\def\prl #1 #2 #3 {{\it Phys.~Rev.~Lett.} {\bf#1} (#2) #3}
\def\mpl #1 #2 #3 {{\it Mod.~Phys.~Lett.} {\bf#1} (#2) #3}
\def\rmp #1 #2 #3 {{\it Rev. Mod. Phys.} {\bf#1} (#2) #3}
\def\sjnp #1 #2 #3 {{\it Sov. J. Nucl. Phys.} {\bf#1} (#2) #3}
\def\cpc #1 #2 #3 {{\it Comp. Phys. Comm.} {\bf#1} (#2) #3}
\def\xx #1 #2 #3 {{\bf#1}, (#2) #3}
\def\preprint{{\it preprint}}

\begin{flushright}
{RAL-TR-1999-009}\\
{January 1999}\\
\end{flushright}

\vspace*{\fill}

\begin{center}
{\Large \bf
The process $e^+e^-\ar H t\bar t$ and its backgrounds\\[0.25cm]
at future electron-positron colliders\footnote{Work supported by 
the UK PPARC.

~$\dagger$ Electronic mail: 
moretti@v2.rl.ac.uk.}}\\[1.0 cm]
{\large S. Moretti$^\dagger$}\\[0.75 cm]
{\it Rutherford Appleton Laboratory,}\\
{\it Chilton, Didcot, Oxon OX11 0QX, UK.}\\[0.5cm]
\end{center}
\vspace*{\fill}

\begin{abstract}
{\noindent 
\small
The process $e^+e^-\ar H t\bar t$ can be used 
at the Next Linear Collider to measure the Higgs-top Yukawa coupling.
In this paper, we compute  $2\ar 8$ processes of the form 
$e^+e^-$ $\ar$ $b\bar b b\bar b W^+W^-$ $\ar$ 
$b\bar b b\bar b \ell^\pm\nu_\ell$ $q\bar q'$, 
accounting for the Higgs-top-antitop signal as well as several 
irreducible backgrounds in the semi-leptonic top-antitop decay channel.
We restrict ourselves to the case of a light Higgs boson in the
range 100 GeV $\Ord$ $M_H$ $\Ord$ 140 GeV.
We use helicity amplitude techniques to compute exactly such
processes at tree level in the framework of the Standard Model.
Total rates and differential spectra of phenomenological interest
are given and discussed.}
\end{abstract}

\vspace*{\fill}
\newpage

\begin{figure}[h!]
~\hskip1.0cm\epsfig{file=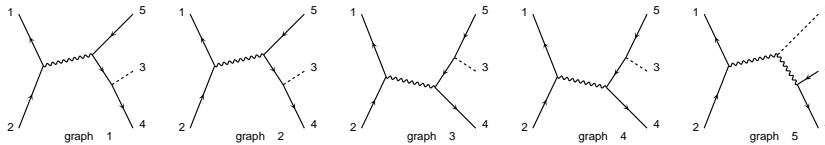,height=12cm}  
\vskip-10.0cm
\caption{Relevant Feynman diagrams contributing at lowest
order to the process $e^+_1 e^-_2 \ar H_3 t_4 \bar t_5$. 
An internal wavy line
represents a $\gamma$ (graphs 1, 3) or a $Z$ (graphs 2, 4, 5).}
\label{fig:eehtt}
\end{figure}

At the Next Linear Collider (NLC), running with a centre-of-mass (CM) energy of
$\sqrt s = 500 $ GeV \cite{ee500}, the Higgs boson
 of the Standard Model (SM)
can be produced in association with top-antitop pairs \cite{eetth}, through 
the process $e^+e^- \ar H t\bar t$, which proceeds via the diagrams
displayed in Fig.~\ref{fig:eehtt}.
That is, the scalar particle can be radiated either from the top quark
pair or from a virtual $Z$ boson. In the latter case, it is the
neutral gauge vector to eventually
produce the heavy quark pair. Clearly, given the actual value of the top 
mass, $m_t\approx175$ GeV, 
between the two sets of graphs, it is the first one which
dominates. On the one hand, the $Z^*\ar t\bar t$ decay occurs far off the
mass-shell of the $Z$ boson. On the other hand, the large Yukawa coupling 
exceeds the strength of the $HZZ$ vertex. Indeed, it is the possibility
of measuring such Yukawa interaction that renders associated production
of Higgs bosons 
and top (anti)quarks phenomenologically interesting at the NLC \cite{top}.  

From the above values of $\sqrt s$ and $m_t$, it follows that
only Higgs scalars with mass $M_H$ up to 140 GeV or so can be produced, because
of the kinematical limit imposed by the difference $\sqrt s-2m_t$. 
For such values of $M_H$, the dominant Higgs decay mode
is $H\ar b\bar b$, this being overtaken by the off-shell
decay into two $W^\pm$'s, i.e., $H\ar W^{+*}W^{-*}$, only for $M_H\OOrd130-140$
GeV, see Fig.~1 of  Ref.~\cite{WJSZK}.
However, these Higgs masses are extremely close
to the kinematical limit of the $H t\bar t$ intermediate state, so that the 
 production cross section of the latter is very small \cite{topthe}. 
Furthermore, notice that in order to
reconstruct the Higgs mass one would require a fully hadronic decay
of the $W^{+*}W^{-*}$ pairs produced in the Higgs 
decay, this leading to a signature with at least eight jets
in the final state. In fact, at least one top quark would be required to
decay into jets, in order to exploit the reconstruction of its mass
to reduce various QCD backgrounds. In other terms, the search for
$H\ar W^{+*}W^{-*}$ decays from $e^+e^-\ar Ht\bar t$ 
would be of difficult experimental use, considering the reduced
number of events, the rather chaotic topology and the problem that
the latter
 generates, because of the combinatorics, while attempting to disentangle
the $H$  and $t$ resonances. In the end, one would be much better off
to rely on the two-body mode $H\ar b\bar b$ over the entire $M_H$ range
allowed by Higgs-top-antitop intermediate states at $\sqrt s=500$ GeV. 

As for $t\bar t$ decays,  one would 
most likely exploit the semi-leptonic channel, i.e.,
$t\bar t \ar b\bar b W^+W^-\ar b\bar b \ell^\pm\nu_\ell$ $q\bar q'$, where
$\ell$ and $\nu$ represent a lepton at high transverse momentum
(to be used for triggering purposes) and its companion neutrino
and $q\bar q'$ refers to the two possible combinations of light quark
pairs and their charge conjugated channels (neglecting 
Cabibbo-Kobayashi-Maskawa mixing effects). 
This is the decay signature we will  concentrate on. As a matter of fact, 
such a choice is not restrictive, in the sense that the latter is to date the 
experimentally preferred channel in searching
for $t\bar t\ar b\bar b W^+W^-$ events \cite{topexp}. 

If one does assume such Higgs and top decay modes, then
signal  events can be searched for in data samples 
made up by four $b$ quark jets, two light quark jets,
a lepton and a neutrino. In other terms, 
a `$4b + 2~{\mathrm{jets}}~+ \ell^\pm + E_{\mathrm{miss}}$' signal,
assuming the four heavy quark jets to be recognised as such thanks to the
$\mu$-vertex devices of the NLC, with $\ell=e,\mu,\tau$\footnote{We include
$\tau$'s to enhance the signal rate, assuming that they
are distinguishable from quark jets.} and where
the missing energy, $E_{\mathrm{miss}}$, originates from
the neutrino escaping detection.

Though the calculation of the on-shell production $e^+e^-\ar Ht\bar t$
has been tackled long ago \cite{first},  that of the complete 
$2\ar 8$ body reaction, without any factorisation of production
and decay processes, has never been attempted before. Not surprisingly
so, as even in presence of only five diagrams, both the large number
of particles in
 the final state and the complicate resonant structure of the latter 
impose non-trivial
problems to the matrix element (ME) calculation and to its integration over 
the phase space, respectively. Things become even more involved
if one starts including (irreducible) backgrounds in the calculation,
as needed in order to realistically simulate phenomenological studies. 
For example, if one restricts oneself to all
those channels that proceed through an intermediate $H b\bar b W^+ W^-$
stage, then the full gauge invariant set  
(including Higgs bosons produced via other graphs than
those in Fig.~\ref{fig:eehtt}) 
counts 350 tree-level diagrams.

\begin{figure}[h!]
~\hskip0.0cm\epsfig{file=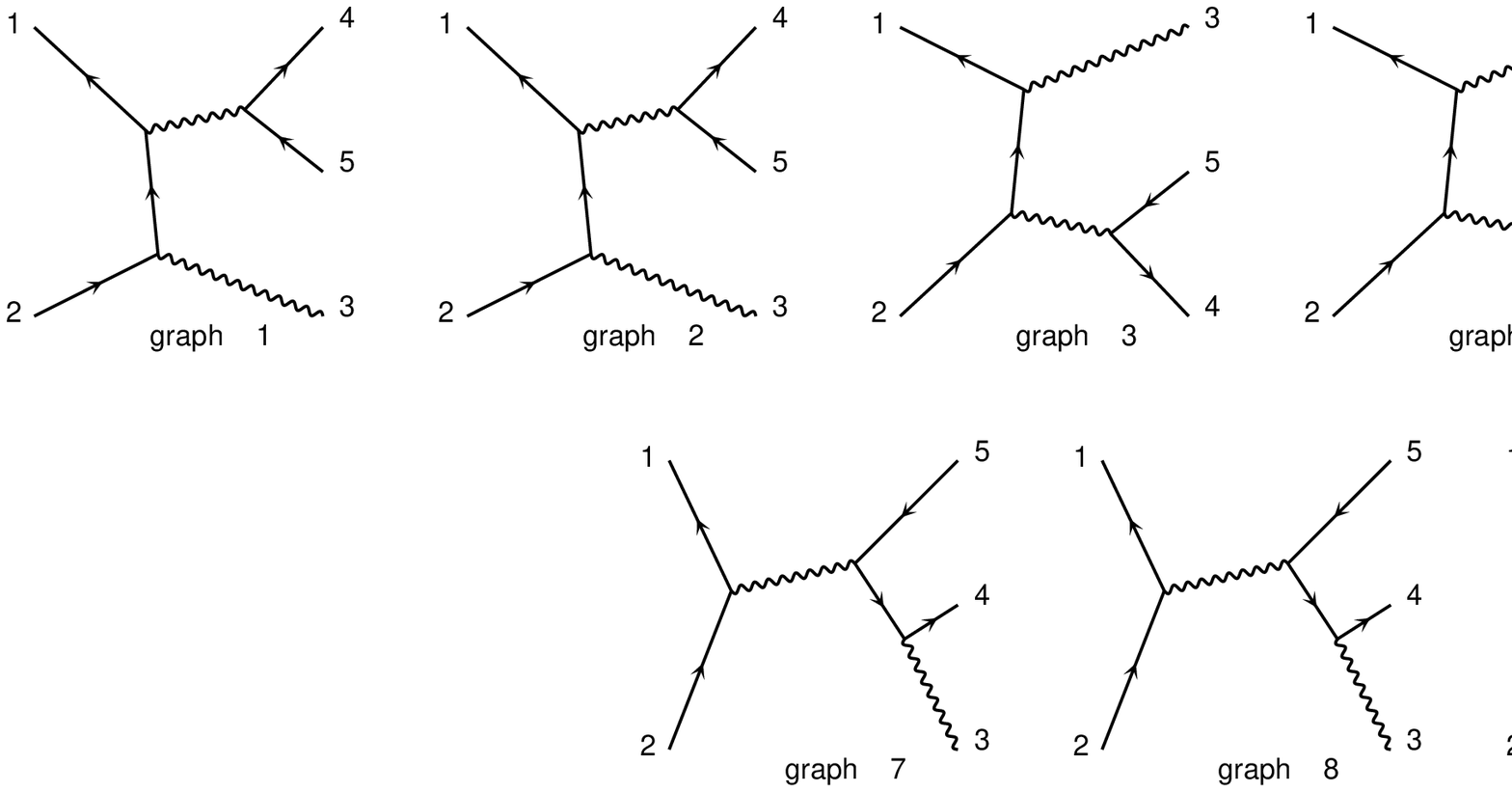,height=12cm}  
\vskip-7.75cm
\caption{Relevant Feynman diagrams contributing at lowest
order to the process $e^+_1 e^-_2 \ar Z_3 t_4 \bar t_5$. 
An internal wavy line
represents a $\gamma$ (graphs 1, 3, 5, 7) or a $Z$ 
(graphs 2, 4, 6, 8, 9).}
\label{fig:eeztt}
\end{figure}

One of the main irreducible backgrounds to the Higgs-top-antitop signal
at the NLC is the scattering $e^+e^-\ar Z t\bar t$ \cite{eettz}, 
if one considers that the two processes
 have comparable production cross sections \cite{eetth,eettz} and that
the $Z$ boson decays into $b\bar b$ pairs some 15\% of the times. Even though
the difference between $M_H$ and $M_Z$ is always larger than 10 GeV
(assuming a late 100 GeV bound on the former from the all of
LEP2 data \cite{LEP})
and the width of the Higgs boson is very narrow (about ten MeV at the most 
for masses up to 140 GeV, see Fig.~2 of  Ref.~\cite{WJSZK}), 
one should recall both the large value of that
of the $Z$ boson, $\Gamma_Z\approx2.5$ GeV, the finite efficiency 
of the detectors in reconstructing jet energies and directions
(to say the least, yielding a resolution of some 5 GeV in invariant mass)
and the mis-assignment problems arising when pairing the four $b$ jets in the
final state in the attempt to recognise resonances in the
$b\bar b$ decay channel. Thus, it is inevitable to conclude 
that $Zt\bar t$ events will represent a serious noise. 
On-shell $Z$-top-antitop production proceeds at tree-level through the
nine graphs of Fig.~\ref{fig:eeztt}. If one however considers, on the
same footing as was done for Higgs production, all the gauge invariant
set of amplitudes producing $ Z b\bar b W^+ W^-$ intermediate states,
followed by $Z\ar b\bar b$,
then the number of graphs involved is 546. (Notice that several of the
production channels described by the latter do involve Higgs bosons, some
of which decay into $b\bar b$ pairs.) 

\begin{figure}[h!]
~\hskip2.0cm\epsfig{file=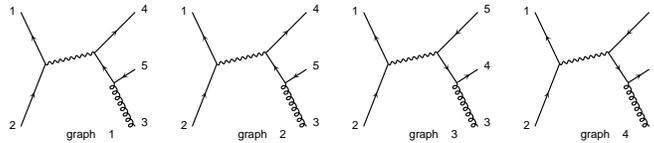,height=12cm}  
\vskip-10.0cm
\caption{Relevant Feynman diagrams contributing at lowest
order to the process $e^+_1 e^-_2 \ar g_3 t_4 \bar t_5$. 
An internal wavy line
represents a $\gamma$ (graphs 1, 3) or a $Z$ (graphs 2, 4).}
\label{fig:eegtt}
\end{figure}

In addition, one should also consider 
$e^+e^-\ar g b\bar b W^+ W^-$ intermediate
states, where $g$ represents a gluon eventually yielding $b\bar b$
pairs. Although none of $b\bar b$ invariant masses has in this
case the tendency of being produced around $M_H$ (in particular, the one 
induced by the $g$ splitting logarithmically increases at very low mass
values, because of the infrared singularity of QCD, only regulated by 
the $b$ mass, $m_b$), such mechanisms proceed through strong interactions,
so that their production rates could well be comparable to those 
of the signal\footnote{For opposite reasons, one can avoid studying 
$e^+e^-\ar \gamma b\bar b W^+ W^-$ reactions, with the photon splitting into
$b\bar b$ pairs.}. 
In fact, because of the mis-pairings of $b$ quarks,
large tails in the $b\bar b$ invariant mass distributions could arise,
despite of the softness and collinearity of two of the
heavy quarks. The dominant background contribution from these
mechanisms would come from $e^+e^-\ar gt\bar t $ events  \cite{eettg}, 
with the gluon radiated before the  (anti)top decays take place. There are four
tree-level diagrams associated with this $2\ar3$
 process, see Fig.~\ref{fig:eegtt}.
The total number of those yielding $g b\bar b W^+ W^-$ states
 is instead 152.

It is the purpose of this letter to compute all such processes
and compare the signal rates and distributions to those obtained
from  the various backgrounds that we have described, in order to
assess the  chances of genuinely exploiting the Higgs-top-antitop production
process in measuring the Higgs-top Yukawa coupling. In this
respect, the reader should notice one subtlety. In fact,
the mentioned coupling not only appears in the 
$e^+e^- Ht\bar t\ar H b\bar b W^+W^-$ `signal', but
also in several `background' mechanisms, such as in $e^+e^-\ar W^{\pm*} W^\mp$
production with one of bosons off-shell, followed by $W^{\pm*} \ar 
H t\bar b + H \bar t b\ar H b\bar b W^\pm$ and $e^+e^-\ar t^*\bar b W^- $
production of an off-shell $t$ quark, eventually yielding $t^*\ar Ht\ar
H b W^+$ (plus the charged conjugate case). These can be regarded
as `single top' processes, as opposed to the `double top' one, i.e.,
$e^+e^-\ar H t\bar t$, themselves being proportional
to the Higgs-top Yukawa coupling. More correctly then, these
two subprocesses should be considered as additional contributions
to the, say, `Yukawa' signal,
further recalling that they carry one resonant top decay (we are 
selecting the semi-leptonic channel, thus implicitly assuming that 
no more than one top mass can in principle be reconstructed). 



To compute all signal\footnote{Note that 
we calculate the Higgs-top-antitop signal at 
the leading-order (LO), though we
are aware that several higher order corrections (mainly 
to the on-shell production) are known to date \cite{topthe,NLO}.
We do this for consistency, as all the $2\ar 8$ background processes 
are evaluated here at tree level.} 
 and background 
graphs we have resorted to helicity
amplitudes methods. In particular, we have made use of the 
{\tt HELAS} subroutines \cite{HELAS}, based on the 
formalism of Ref.~\cite{HZ}. All the {\tt FORTRAN} codes produced this way
have been tested for gauge invariance satisfactorily,
 so to give us confidence in our numerical 
results. Furthermore,
 the $2\ar8$ `dominant' (as we shall see below)
signal and background 
processes of the form $e^+e^-\ar X b\bar b W^+W^-\ar
b\bar b b\bar b \ell^\pm\nu_\ell q\bar q'$, with $X=H,Z$ and $g$, 
have also  been implemented by using the spinor techniques described
in Refs.~\cite{Berends,ioPRD}. 
Wherever the two approaches overlapped, we have
seen perfect agreement between the outputs of the
two sets of codes.

Numerical results have been produced after integration of the 
Feynman amplitudes squared over eight-body phase spaces. In order to 
account accurately for all their components, we have split the 
MEs of the form $e^+e^-\ar Xb\bar b W^+W^-\ar
b\bar b b\bar b \ell^\pm\nu_\ell q\bar q'$ 
in resonant sub-terms and integrated each of these separately.
Only in the end the various integrals were
summed up, in order to recover gauge-invariance \cite{paps}.  
The algorithms used to perform the multi-dimensional integrations
were {\tt VEGAS} \cite{VEGAS}
and, for comparison, {\tt RAMBO} \cite{RAMBO}.

To describe the vector and axial couplings
of the gauge bosons to the fermions, we have used $\sin^2\theta_W=0.2320$.
The strong coupling constant $\alpha_s$ entering the QCD processes 
(i.e., $X=g$) has been evaluated
at two loops, with $N_f=4$ and 
$\Lambda_{\overline{\mathrm {MS}}}=230$ MeV,
at a scale equal to the collider CM energy, $\sqrt s=E_{\mathrm{ecm}}=500$
GeV. The electromagnetic coupling was  $\alpha_{em}=1/128$.
For masses and widths, we have used:
$m_\ell=m_{\nu_\ell}=m_u=m_d=m_s=m_c=0$,
$m_b=4.25~{\mathrm{GeV}}, m_t=175~{\mathrm{GeV}}$  (as default),
$M_Z=91.19~{\mathrm {GeV}}, \Gamma_Z=2.50~{\mathrm {GeV}}$,
$M_W=80.23~{\mathrm {GeV}}$ and $\Gamma_W=2.08~{\mathrm {GeV}}$.
As for the top width $\Gamma_t$, 
we have used the LO value of 1.5 GeV.
Only in one circumstance, in order to study the sensitivity of the signal
processes to the Higgs-top Yukawa coupling,
we have changed $m_t$ by $\pm5$ GeV. The widths
corresponding to these two new values are 1.3 and 1.6, for the lower
and higher $m_t$ figure, respectively.

Concerning the Higgs boson, we have spanned its mass $M_H$ over the
range 100 to 140 GeV. As for its width, $\Gamma_H$, 
we have computed it by means of the same program
described in Ref.~\cite{WJSZK}, which  uses 
a running $b$ mass in evaluating the $H\ar b\bar b$ decay fraction. Thus,
for consistency, we have evolved here the value of $m_b$ entering the
$Hbb$ Yukawa coupling of the $H\ar b\bar b$ decay current  
in the same way as then. 

Finally, notice that starting from our $2\ar8$ MEs for
$e^+e^-\ar X t\bar t\ar b\bar b b\bar b W^+W^-\ar
b\bar b b\bar b \ell^\pm\nu_\ell q\bar q'$, in all cases $X=H,Z$ and $g$,
we are able to reproduce (apart from minor spin correlations) 
the cross sections
that one obtains from the $2\ar 3$ ones for $e^+e^-\ar X t\bar t$,
times the relevant branching ratios (BRs), by
adopting a Narrow Width
Approximation (NWA) for the various resonances $R$ involved (i.e., 
$R=H$, $t$, $W^\pm$ and $Z$),  by rewriting the corresponding
(denominator of the) propagators as (for $\Gamma\equiv\Gamma_R$
the standard expression is recovered): 
\begin{equation}\label{NWA}
\frac{1}{p^2-m_R^2+{\mathrm{i}}m_R\Gamma}
\left(\frac{\Gamma}{\Gamma_R}\right)^{1/2},
\end{equation}
with $\Gamma\ar0$, 
this way mimicking a delta distribution,
i.e., $\delta(p^2-m_R^2)$.
(In the case $X=g$ we had to supplemented the $2\ar3$ ME for 
$e^+e^-\ar gt\bar t$ with the splitting function for
$g\ar b\bar b$.)
  
In the following, total and differential rates are those 
at parton level, as we identify
jets with the partons from which they originate. Gaussian smearing effects are
simulated. No efficiency  to tag four $b$ quarks is included.


We start our analysis of the results with a disclaimer: we have not included
Initial State Radiation
(ISR) \cite{ISR} in our calculations. 
We have done so mainly for technical reasons.
Simply because we are already dealing with complicated processes requiring
delicate integrations, over nineteen dimensions and with
a laborious rearrangement of the phase space, to account for the
 multi-resonant behaviour of hundred of diagrams, that even
adding the ISR in the simplest way\footnote{For example, via the
so-called Electron Structure Function (ESF) approach \cite{ISR}.}
would prove rather costly in terms of efficiency of the computation.
In addition, we would expect ISR to affect rather similarly the
various processes of the form $e^+e^-\ar X b\bar b W^+W^-\ar b\bar b
b\bar b \ell^\pm\nu_\ell q\bar q'$. As we are basically interested
in relative rates among the latter, we are confident that the basic
features of our results are indifferent to the presence or not
of photons radiated by the incoming electron-positron beams\footnote{We
also neglect beamsstrahlung and Linac energy
spread, by assuming a narrow beam design \cite{ISR}.}. 

\begin{figure}[b]
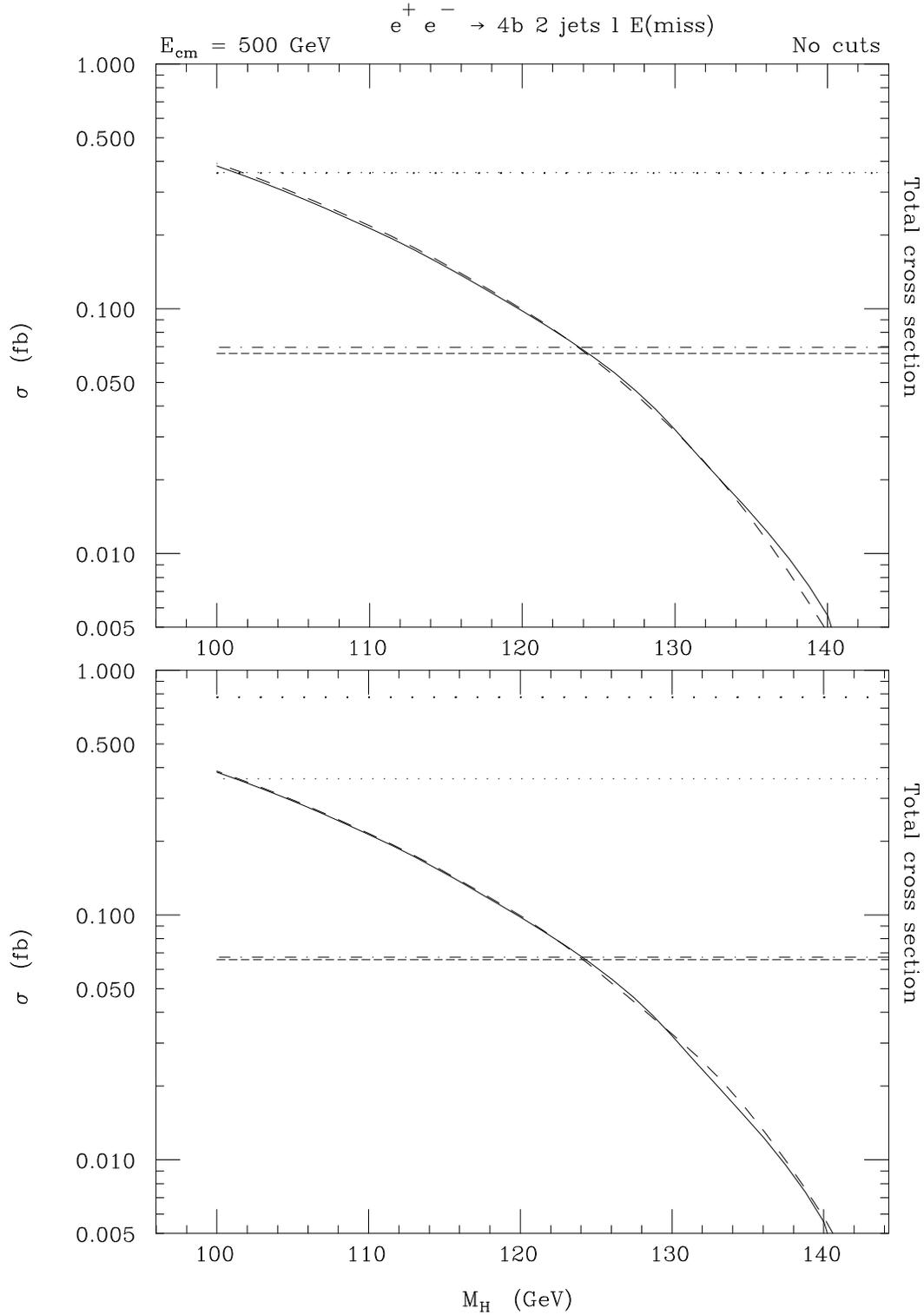

~\epsfig{file=cross1.ps,height=14cm,angle=90}
\vskip+0.0005cm
~\epsfig{file=cross2.ps,height=14cm,angle=90}
\caption{Cross sections for the following processes (see the text): 
(\ref{htt}) (long-dashed line, above), 
(\ref{ztt}) (dot-dashed line, above) and 
(\ref{gtt}) (dotted line, above); 
(\ref{htt_hbbww}) (solid line, above and below), 
(\ref{ztt_zbbww}) (dashed line, above and below) and 
(\ref{gtt_gbbww}) (fine-dotted line, above and below); 
(\ref{hbbww}) (long-dashed line, below), 
(\ref{zbbww}) (dot-dashed line, below) and 
(\ref{gbbww}) (dotted line, below). 
The total CM energy is $\protect{\sqrt s\equiv E_{\mathrm{cm}}=500}$ GeV.
No cuts have been implemented.}
\label{fig:cross}
\end{figure}

Fig.~\ref{fig:cross} presents the production cross sections
 for the following (sub)processes:
\begin{enumerate}
\item firstly, the $2\ar 3$ on-shell ones,
\begin{equation}\label{htt}
e^+e^-\ar H t\bar t,
\end{equation}
\begin{equation}\label{ztt}
e^+e^-\ar Z t\bar t,
\end{equation}
\begin{equation}\label{gtt}
e^+e^-\ar g t\bar t,
\end{equation}
as obtained from the diagrams in Fig.~\ref{fig:eehtt}--\ref{fig:eegtt}  
multiplied by the  
BRs and the $g\ar b\bar b$ splitting function;
\item secondly, the $2\ar 8$ ones which proceed via those above,
\begin{equation}\label{htt_hbbww}
e^+e^-\ar H t\bar t\ar H b\bar b W^+W^-\ar b\bar b b\bar b \ell^\pm
\nu_\ell q\bar q',
\end{equation}
\begin{equation}\label{ztt_zbbww}
e^+e^-\ar Z t\bar t\ar Z b\bar b W^+W^-\ar b\bar b b\bar b \ell^\pm
\nu_\ell q\bar q',
\end{equation}
\begin{equation}\label{gtt_gbbww}
e^+e^-\ar g t\bar t\ar g b\bar b W^+W^-\ar b\bar b b\bar b \ell^\pm
\nu_\ell q\bar q',
\end{equation}
as obtained from the diagrams in Fig.~\ref{fig:eehtt}--\ref{fig:eegtt} 
supplemented with the  decay currents;
\item thirdly, the $2\ar 8$ ones including also all other diagrams,
\begin{equation}\label{hbbww}
e^+e^-\ar b\bar b W^+W^- \ar b\bar b b\bar b \ell^\pm
\nu_\ell q\bar q',
\end{equation}
\begin{equation}\label{zbbww}
e^+e^-\ar Z b\bar b W^+W^- \ar b\bar b b\bar b \ell^\pm
\nu_\ell q\bar q',
\end{equation}
\begin{equation}\label{gbbww}
e^+e^-\ar g b\bar b W^+W^- \ar b\bar b b\bar b \ell^\pm
\nu_\ell q\bar q'.
\end{equation}
\end{enumerate}

In moving from cases 1. to 2., one can appreciate the onset of spin
and width effects, see top of Fig.~\ref{fig:cross}, whereas in comparing 
2. and 3. one can disentangle
those due to the diagrams not proceeding via $X$-top-antitop pairs,
see bottom of Fig.~\ref{fig:cross}. It turns out that  spin and width
effects are sizable only for the Higgs-top-antitop and $Z$-top-antitop
processes, not for the gluon-top-antitop ones. They are of the order of $+6\%$
in $Zt\bar t$ diagrams, whereas in the case of $Ht\bar t$ they vary between
$+2.5\%$ at $M_H=100$ GeV and $-15\%$ at $M_H=140$ GeV. For $gt\bar t$ diagrams
they amount to less than $1\%$ (hence the
overlapping of the two dotted curves in the top frame of
Fig.~\ref{fig:cross}). As for effects due to 
non-$X$-top-antitop graphs, things go the 
other way around. The $gt\bar t$ rates
are hugely increased, 
by as much as a factor of two, whereas the $Zt\bar t$ and $Ht\bar t$ ones
never get larger than $2.3\%$ and $4.3\%$, respectively. The growth of the 
QCD rates is mainly 
due to the large amount of gluon radiation (here, eventually
yielding $b\bar b$ pairs) produced 
in the top quark decays \cite{eettg}. Note, however,
that the latter can easily be controlled by imposing that none of
the invariant masses of five particle systems with three heavy and two light 
quarks (and/or two leptons, if the $\nu_\ell$ momentum is reconstructed) 
reproduces $m_t$.

Therefore,
by studying the production rates of all reactions (\ref{htt})--(\ref{gbbww}), 
one may remark on two key aspects. On the one hand, 
the bulk of the cross sections of processes
(\ref{hbbww})--(\ref{gbbww}) comes  from the $X$-top-antitop channels 
(\ref{htt_hbbww})--(\ref{gtt_gbbww}). On the other hand, the QCD process
(\ref{gtt_gbbww})
is the dominant one, for any value of $M_H$. (That for $M_H\OOrd125$
GeV or so the $Zt\bar t$ production rates started exceeding the
$Ht\bar t$ ones was rather trivial to derive \cite{eetth,eettz}.) 
Whereas the first result was
clearly expected, the second one came as somewhat of a surprise.
As a consequence, in the reminder of our analysis, we will mainly concentrate
on the $X$-top-antitop diagrams and study some of their differential spectra
that can help disentangling the Higgs diagrams from the $Z$ and, especially,
the gluon ones. We will do so for the choice $M_H=130$ GeV, as representative
of the case in which both $Z$-top-antitop and
gluon-top-antitop backgrounds overwhelm the Higgs-top-antitop
signal (see Fig.~\ref{fig:cross}).

\begin{figure}[!t]
\begin{center}
\vskip-3.0cm
~\epsfig{file=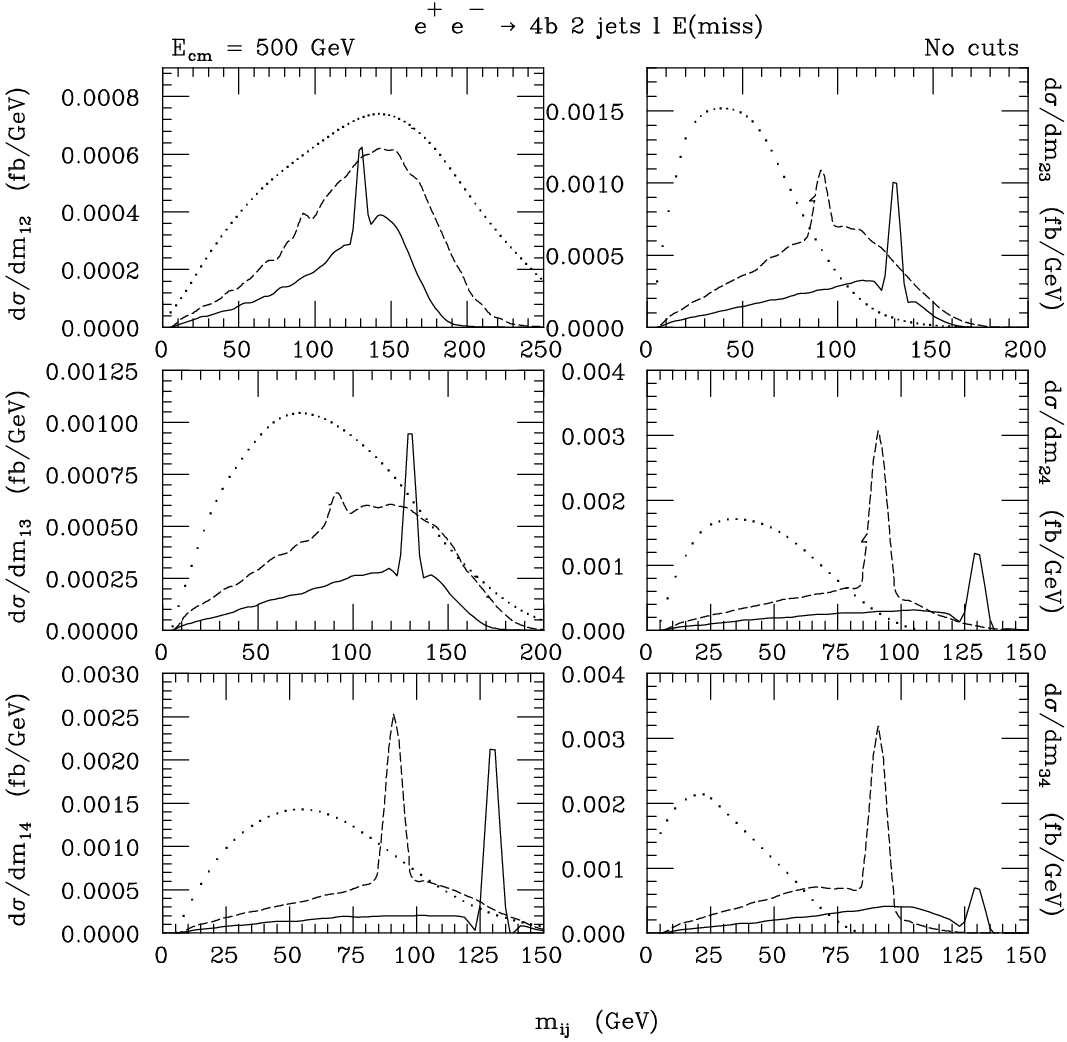,width=16cm,height=18cm,angle=0}
\caption{Differential distributions in the invariant mass
of all possible combinations $ij$, with $i<j=2, ... 4$,  of the
energy-ordered $b$ jets (i.e., such that $E_1>E_2>E_3>E_4$)  for the 
following processes (see the text):
(\ref{htt_hbbww}) (solid line), 
(\ref{ztt_zbbww}) (dashed line) and 
(\ref{gtt_gbbww}) (dotted line). 
The total CM energy is $\protect{\sqrt s\equiv E_{\mathrm{cm}}=500}$ GeV.
No cuts have been 
implemented. Note that the rates of reaction (\ref{gtt_gbbww}) 
have been divided by three for readability. 
}
\label{fig:mbb}
\end{center}
\end{figure}

As we have already stressed that one of the $b\bar b$ pairs in the final
state would naturally resonate at $M_H$, at $M_Z$ or logarithmically
increase at low mass, for processes
(\ref{htt_hbbww}), (\ref{ztt_zbbww}) and (\ref{gtt_gbbww}), respectively,
we start investigating the di-jet mass spectra that can be reconstructed from
the four $b$ quarks in
the `$4b + 2~{\mathrm{jets}}~+ \ell^\pm + E_{\mathrm{miss}}$' signature.
Since we do not assume any jet-charge determination of 
the ($\mu$-vertex tagged) $b$ jets and consider  negligible the mis-tagging 
of light-quark jets as heavy ones, six such combinations can be built up.
We distinguish among these by ordering the four $b$ jets in energy
(i.e., $E_1>E_2>E_3>E_4$),  in such a way that 
the $2b$ invariant mass $m_{ij}$ refers to the $ij$ pair (with
$i<j=2,3,4$) in which the $i$-th and $j$-th most energetic particles enter.
Having done so, one should expect to see the typical resonant/logarithmic 
behaviours
described above now `diluted' in the various $ij$ combinations. This is
evident from  Fig.~\ref{fig:mbb}. There, one can appreciate the
resonant shapes around $M_H$ and $M_Z$ in all $ij$ cases (for $ij=12$, 
the $Z$ peak is just a tiny kink on top of a Jacobian shape).
As for the `divergence' in the $g\ar b\bar b$ splitting of the QCD
process, this can easily be spotted in the case $ij=34$. In the end, the $2b$
mass spectra look rather promising as a mean of reducing
 both backgrounds (\ref{ztt_zbbww})--(\ref{gtt_gbbww}).
By requiring, e.g., $m_{34}>50$ GeV, one would vigorously reduce
the latter; similarly, by imposing, e.g., $|m_{14}-M_Z|>15$ GeV one would
reject the former considerably.

\begin{figure}[!t]
\begin{center}
\vskip-3.0cm
~\epsfig{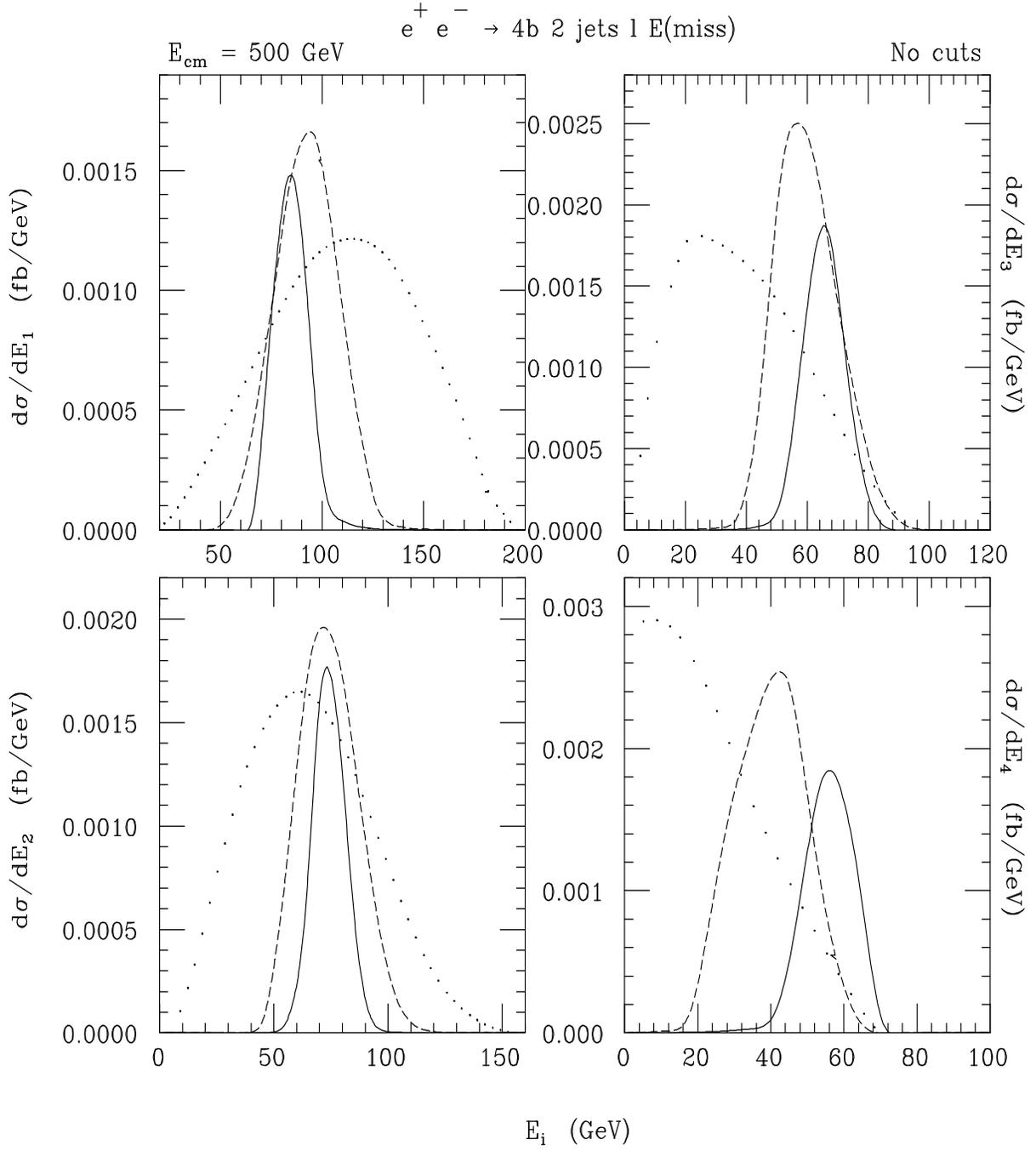}
\caption{Differential distributions in energy of the
energy-ordered $b$ jets (i.e., such that $E_1>E_2>E_3>E_4$)  for the 
following processes (see the text):
(\ref{htt_hbbww}) (solid line), 
(\ref{ztt_zbbww}) (dashed line) and 
(\ref{gtt_gbbww}) (dotted line). 
The total CM energy is $\protect{\sqrt s\equiv E_{\mathrm{cm}}=500}$ GeV.
No cuts have been 
implemented. Note that the rates of reaction (\ref{gtt_gbbww}) 
have been divided by three for readability. 
}
\label{fig:Eb}
\end{center}
\end{figure}

Another way of looking at the same phenomenology in
processes (\ref{htt_hbbww})--(\ref{gtt_gbbww}) is by studying the
energy spectra of the four $b$ quarks. In fact, the larger value of
$M_H$, as compared to $M_Z$, should boost the $b$ quarks generated by
the Higgs boson towards energies higher than those
achieved in the $Z$ decays. Conversely, the energy of the $b$ quarks
emerging from the two remaining unstable particles, top and antitop
quarks, should be softer in the first case. Following similar arguments,
one should expect the hardest(softest) $b$ (anti)quark from gluon events to
actually be the hardest(softest) of all cases
(\ref{htt_hbbww})--(\ref{gtt_gbbww}), once again, because of the
infrared QCD splitting of a soft gluon. Recalling that the two
most energetic $b$'s seldom come from a $H$, $Z$ or $g$ splitting in
$X b\bar b W^+W^-$ intermediate states (see top-left 
curves in Fig.~\ref{fig:mbb}), the above kinematic features are clearly 
recognisable in Fig.~\ref{fig:Eb}. Therefore, the energy spectra
too are  rather useful in disentangling Higgs events.
If one imposes, e.g., 
$E_1<100$ GeV and $E_4>50$ GeV, both $Z$ and gluon events
can be strongly depleted, at a rather low cost for the signal.

We conclude the numerical analysis
 by studying the sensitivity of the signal to the 
Higgs-top Yukawa coupling, by varying the top mass by 5 GeV above and below 
its default value. However, as to modify $m_t$ (and, consequently, $\Gamma_t$)
has also incidence on the top propagators, and since these enter many of the
diagrams associated with processes (\ref{hbbww})--(\ref{gbbww}), we
present the rates for the latter, that is, for the full sets
of diagrams in each case. This is done in Tab.~\ref{tab:yukawa}. The value 
we have chosen for the Higgs mass, i.e., $M_H=130$ GeV, is a critical one
for process (\ref{htt_hbbww}), the main source of events (\ref{hbbww}).
In the sense that the sum $2m_t+M_H$ is very close to $\sqrt s$, so
that the corresponding rates in Tab.~\ref{tab:yukawa} 
(see second column) are the result
of the interplay between the rise of the cross section with $m_t^2$
and its fall because of the phase space suppression (width effects
are less relevant). Indeed, between the two
tendencies is the latter to dominate. In fact, the production
cross sections of all three processes (\ref{hbbww})--(\ref{gbbww})
decrease with increasing top mass.
Even in presence of such delicate interplay, the sensitivity
of Higgs-top-antitop events
to the actual value of the top mass is rather strong, as the corresponding 
cross section changes by a factor of 5 between $m_t=170$ and 180 GeV.  
Backgrounds variations are always smaller. However, both processes
(\ref{zbbww})--(\ref{gbbww}) are larger than (\ref{hbbww}).
Once again, it has to be stressed that
background rates ought to be reduced severely if one
wants to perform dedicated studies of the Higgs-top Yukawa coupling.   

\begin{table}[t!]
\begin{center}
\vskip-0.5cm
\begin{tabular}{|c||c|c|c|}

\hline


\multicolumn{4}{|c|}
{$\sigma_{\mathrm{tot}}$ (fb)}
\\ \hline\hline

$m_t$ (GeV)                          & 
$H b\bar b W^+W^-$                   & 
$Z b\bar b W^+W^-$                   &     
$g b\bar b W^+W^-$                  \\ 
\hline
170                              &
$0.061$                          & 
$0.087$                          & 
$0.79$                          \\
\hline
175                              &
$0.033$                          & 
$0.067$                          & 
$0.77$                          \\
\hline
180                              &
$0.012$                          & 
$0.050$                          & 
$0.74$                          \\
\hline


\end{tabular}
\end{center}
\vskip-0.5cm
\caption{Cross sections for processes (see the text)
(\ref{hbbww})--(\ref{gbbww}), for three discrete values of the top mass.
The total CM energy is $\protect{\sqrt s\equiv E_{\mathrm{cm}}=500}$ GeV.
The Higgs boson mass is $M_H=130$ GeV.
No cuts have been implemented. (Numerical errors from the Monte Carlo
 integration do not affect the significant digits shown.)}
\label{tab:yukawa}
\vskip-0.5cm
\end{table}


In summary, in our opinion, the study of the Higgs-top Yukawa coupling
at future electron-positron colliders, such as the NLC running with
a CM energy of 500 GeV, can in principle be pursued by means of the 
Higgs-strahlung process $e^+e^-\ar Ht\bar t$.
In fact, the irreducible backgrounds affecting the
latter can be brought under control in the semi-leptonic top-antitop 
decay channel 
$t\bar t\ar b\bar b W^+W^-\ar b\bar b \ell^\pm\nu_\ell q\bar q'$, further
assuming $H\ar b\bar b$, as natural for Higgs masses up to 140 GeV or so.

However, this requires to somehow recognise the $b$ jets in the final state 
with high efficiency, as the observable  rates of the signal are 
below the femtobarn level. The knowledge of the momenta of the heavy quark
jets entering 
the signature `$4b + 2~{\mathrm{jets}}~+ \ell^\pm + E_{\mathrm{miss}}$'
is crucial in order to reduce the overwhelming QCD background, mainly
proceeding via $e^+e^-\ar gt\bar t$ events, if one aims to
disentangle such Higgs events at all. The competing 
electroweak background, mainly proceeding through $e^+e^-\ar Zt\bar t$ 
intermediate states, can be dealt with if the mass resolution of
di-jet pairs of $b$ quarks is around 10 GeV or less.
Other irreducible background channels, induced by $e^+e^- \not\ar Xt\bar t
\ar X b\bar b$
$W^+W^-$ intermediate states, with $X=H,Z$ or $g$, are significantly
smaller than those proceeding via $X$-top-antitop graphs, with the only
exception of QCD graphs involving one radiative (anti)top decay.

In the end then, 
although a careful simulation of possible tagging strategies should
eventually be performed, 
we believe that, if the Higgs mass turns out to be in the intermediate
range, the NLC constitutes an ideal laboratory for the kind of studies
sketched here. We base our conviction on the fact the we have 
performed a new and rather complete calculation of  
signal and backgrounds involving up to ten external particles.

\end{document}